\documentclass[conference, 10pt]{IEEEtran}
\IEEEoverridecommandlockouts
\usepackage{algpseudocode}
\usepackage{algorithm}
\usepackage{amsmath,amssymb,amsfonts}
\usepackage{graphicx}
\usepackage{footnote}
\usepackage{url}
\usepackage{textcomp}
\usepackage[nolist,nohyperlinks]{acronym}
\usepackage{xcolor}
\usepackage{balance}
\usepackage[tight,footnotesize]{subfigure}
\usepackage{multirow}
\def\BibTeX{{\rm B\kern-.05em{\sc i\kern-.025em b}\kern-.08em
    T\kern-.1667em\lower.7ex\hbox{E}\kern-.125emX}}
\begin{document}
\newcommand{\ourmethod}{\textit{SmartMME}\;}
\title{\ourmethod{}: Implementation of Base Station Switching Off Strategy in ns-3}

\author{\IEEEauthorblockN{Argha Sen\IEEEauthorrefmark{1}, Bhupendra Pal\IEEEauthorrefmark{2}, Seemant Achari\IEEEauthorrefmark{3}}
\IEEEauthorblockA{Department of Computer Science and Engineering,
Indian Institute of Technology Kharagpur, India\\
Email: \IEEEauthorrefmark{1}arghasen10@gmail.com,
\IEEEauthorrefmark{2}bipin.s.pal@gmail.com,
\IEEEauthorrefmark{3}seemanthachari@gmail.com,
}
}




\maketitle
\begin{abstract}
In the landscape of next-generation cellular networks, a projected surge of over 12 billion subscriptions foreshadows a considerable upswing in the network's overall energy consumption. The proliferation of \ac{UE} drives this energy demand, urging 5G deployments to seek more energy-efficient methodologies. In this work, we propose \ourmethod{}, as a pivotal solution aimed at optimizing Base Station (BS) energy usage. By harnessing and analyzing critical network states—such as UE connections, data traffic at individual UEs, and other pertinent metrics—our methodology intelligently orchestrates the BS's power states, making informed decisions on when to activate or deactivate the BS. This meticulous approach significantly curtails the network's overall energy consumption. In a bid to validate its efficiency, we seamlessly integrated our module into Network Simulator-3 (ns-3), conducting extensive testing to demonstrate its prowess in effectively managing and reducing net energy consumption. As advocates of collaborative progress, we've opted to open-source this module, inviting the engagement and feedback of the wider research community on GitHub\footnote{\url{https://github.com/ns3-dev/ns3-mmwave} (Accessed: \today)}.          
\end{abstract}

\section{Introduction}
To meet the vast network traffic demand, next-generation cellular networks will deploy a huge number of small-scale \ac{5G} \ac{BS}s. These dense deployments can meet the network traffic demand, but at the same time, they will consume more energy in comparison to the legacy \ac{4G} \ac{LTE} deployments. Since the launch of the first commercial \ac{5G} \ac{NSA} networks, there has been an impact on the battery life of about $10\%$ more battery usage on \ac{5G} in comparison to \ac{4G}~\cite{s_2022}. To enable green communication for a sustainable future, researchers are working on designing energy-efficient techniques for next-generation wireless cellular networks.

Many of the previous works proposed \ac{BS} switching ON/OFF strategies based on the underlying network traffic~\cite{ye2019drag, gao2019user, wu2021deep}. The core idea is to capture the continuous state of traffic demand at the \ac{BS} and, accordingly, take the switching on/off decision for the \ac{BS}. Some of these works formulate the on/off switching problem into a Markov Decision Process that can be solved by some reinforcement learning methods~\cite{ye2019drag,wu2021deep}. These approaches meticulously harness spatial and temporal data traffic correlations to predict traffic surges at BSs, aligning on/off decisions with energy consumption or network performance metrics in a reward-based system.  However, most of these studies have predominantly relied on mathematical simulations to substantiate their energy-saving assertions.

The development of 5G technology is still ongoing and not widely available, especially in middle- and lower-income countries. Thus, to study power-saving schemes in 5G \ac{NR}, some researchers use network simulators like ns-3, which save time and money by allowing them to validate their solutions without needing a physical prototype. The ns3-mmWave module~\cite{mezzavilla2018end} is used to provide an end-to-end 5G protocol stack for \ac{NR} simulations in ns-3, and it has been utilized in various recent research works~\cite{zhang2017ns,zugno2021extending,zugno2019simulation, palit2023improving, sen2023revisiting}. Instead of mathematical simulations, a real network simulator helps capture other network impacts, such as channel delay, propagation loss, etc., that resemble a real network.

In this paper, broadly, we aim to design the energy-aware networking method in a network emulation platform, ns-3, in which the sleep or wakeup switching choices at the \acp{BS} are triggered depending on the end user's QoE needs. The ns-3 implementation will help the research community validate their implementations of \ac{BS} switching off strategies in a real network simulator instead of validating them purely on mathematical simulations. However, this pursuit faces various research and engineering hurdles. \textbf{First}, understanding how an imminent \ac{BS} shutdown in ns-3 might impact QoE for streaming media applications on linked UEs becomes imperative. \textbf{Second}, the decision function for \ac{BS} shutdown should be simple, so the signalling cost for coordination between surrounding \ac{BS} is minimal. \textbf{Third}, since we want to optimize the energy consumption of the whole network, we need to set up energy modules for both the \ac{UE} and the \ac{BS}, accompanied by the need to implement the \ac{BS} switching off functionalities in ns-3. \textbf{Fourth}, we need to connect the switching-off strategy with the \ac{UE} data traffic so that the switching-off doesn't impact the \ac{QoE} of the UE. Conquering these challenges promises a method where \acp{BS} in ns-3 dynamically toggle between sleep and active modes, optimizing the network holistically.

Although ns-3 lacks a well-defined \ac{UE} and \ac{BS} energy modules to implement and test the energy-efficient algorithms for 5G \ac{NR}. With the help of the implementation guidelines provided in~\cite{sen2021ns3}, we have developed the \ac{UE} energy module considering the 3GPP specification $38.840$~\cite{38840} and tested that on the ns3-mmWave~\cite{mezzavilla2018end} framework. Similarly, based on the implementation guidelines in~\cite{sen2022ns3}, our work includes developing an energy module for the \ac{BS}.

Now to develop the \ac{BS} switching off effectively, a key consideration emerges: if we want to enhance the net QoE of the \ac{UE}, then we need to keep the \ac{BS} always powered on, which helps in better energy efficiency at the UE end but leads to higher energy consumption at the \ac{BS} end. On the other hand, if we randomly switch off the 5G \ac{BS}, that may provide better energy efficiency at the \ac{BS} end but may degrade the battery performance of the UE because of repeated re-transmissions. Keeping the dependencies and trade-offs in mind, our focus is an overall network optimization that covers all components' optimization together.

We have developed \ourmethod{}: a \textbf{Smart} \textbf{M}obility \textbf{M}anagement \textbf{E}ntity, designed to capture crucial control channel information encompassing UEs' mobility context, handovers, and connections established with the \ac{BS}s, etc. This entity is instrumental in determining the operational state of a \ac{BS}, evaluating its IDLE status, and facilitating automatic shutdown if deemed idle. It also checks the data traffic at the \ac{BS}. For that purpose, it checks whether the connected UEs are IDLE. If it finds the UE to be IDLE but still connected at the \ac{BS}, it allows the \ac{BS} to get switched off. This functionality has been integrated into ns-3 through the incorporation of supplementary helper functions augmenting the Physical Layer's state-switching mechanisms.

Efficiently powering down an IDLE \ac{BS} is instrumental in conserving considerable energy that would otherwise be needlessly consumed during inactive periods. However, executing this task encounters challenges due to the dynamic and unpredictable movements of UEs. An intriguing avenue involves the development and deployment of a sophisticated predictive model at the \ac{BS}~\cite{zhang2022intelligent}. This model, designed to discern patterns in the frequent or infrequent movements of UEs within specific timeframes, enables automated toggling of the \ac{BS} on/off status. While this predictive framework offers substantial promise, its implementation could introduce intricacies and additional overhead to the switching-off strategy. To refine and streamline this approach, \ourmethod{} takes a nuanced stance by intricately observing the UE mobility context. It continuously monitors and assesses whether UEs have entered zones where \ac{BS} power is deactivated and subsequently activates or deactivates the \ac{BS} based on these insights.

In summary, the contributions of this paper are as follows:
\begin{enumerate}
    \item Development of \ourmethod{}, that captures the control channel information of UEs mobility context, data traffic, etc., and accordingly switches on/off the 5G \ac{BS}s. 
    \item Implemented the \ac{BS} switching off helper functions.
    \item Added energy consumption model at the UE and the \ac{BS} to capture the network's energy consumption.  
    \item Implemented switching-off strategies which are aware of the number of UEs connected to a typical \ac{BS} and also aware of the data traffic in that \ac{BS}.
    \item Finally, we evaluated and tested the module in ns-3 and open-sourced the implementation in GitHub.
\end{enumerate}

The rest of this work is structured as follows. Section~\ref{sec:rel_work} provides brief related works in this domain. Section~\ref{sec:sys_arcgh} describes the methodologies involved in designing \ourmethod{} in ns-3. In Section~\ref{sec:eval}, we evaluate \ourmethod{} and show its performance in lowering the net energy consumption of the network. Finally, in Section~\ref{sec:conclusion}, we conclude with some possible future directions.
\section{Related Work}\label{sec:rel_work}
Existing literature on \ac{BS} switching-off problems branches out into two main directions, (i) Heuristic-based optimization problems which are typically solved using heuristic algorithms, i.e., problem-solving methods that use trial and error to find the solution, and (ii) Markov Decision Problem (MDP) that involve making a sequence of decisions over time, where the outcomes of the decisions are uncertain and depend on the previous decisions made. Here, we summarize the most notable works in each of these directions. 

\subsection{Heuristic based Approach}
Authors in \cite{gao2019user, cai2016green, de2023modelling} considered the \ac{UE} mobility pattern in deciding the \ac{BS} switching off strategy. In \cite{gao2019user} the \ac{BS}s are turned off according to their loads and the amount of time required for mobile users to arrive at a given \ac{BS}. In another heuristic approach~\cite{yu2016minimizing}, authors suggested methods to decrease energy costs by dynamically activating and deactivating \acp{BS} based on switching energy costs and the transmission power adjustment of \acp{BS}. Liu et al.~\cite{liu2015small} proposed an energy-efficient maximization problem based on coverage probability under either a random or strategic sleeping policy.

Ye et al.~\cite{ye2013user} proposed a combined strategy that accomplishes load-aware performance optimization by taking into consideration cell affiliation and resource allocation, as well as including biasing factors such as SINR and rate in their formulation. Also \cite{tang2017energy,huang2018dynamic, abdulkafi2014energy} considered load aware \ac{BS} switching OFF strategies. In yet another study~\cite{huang2018dynamic}, propposed cell association scheme with load balancing scheme and energy efficiency-based dynamic switching strategies, guaranteeing minimum SINR of UEs. \cite{abdulkafi2014energy} studied a small-cell sleep approach based on the traffic load in which BSs are switched off for a given amount of time when their traffic load falls below a particular threshold. 

\begin{table}
\scriptsize
\caption{Previous Works and their approach in the light of simulation platform}
\label{tab:rel_work}
\begin{tabular}{|l|l|l|l|}
\hline
\multicolumn{1}{|c|}{Paper} & \multicolumn{1}{c|}{Objective}                                                                                                                                                            & \multicolumn{1}{c|}{\begin{tabular}[c]{@{}c@{}}Real \\ n/w\end{tabular}} & \multicolumn{1}{c|}{\begin{tabular}[c]{@{}c@{}}Mathematical \\ Simulation\end{tabular}} \\ \hline

\cite{abdulkafi2014energy}                          & \begin{tabular}[c]{@{}l@{}}Traffic load balancing and sleep \\ mode mechanisms\end{tabular}                                                                                                                                           & No & Yes                                                                                     \\ \hline
\cite{gao2019user}                                  & \begin{tabular}[c]{@{}l@{}}n/w loads and the amount of time \\ required for UE to arrive at a BS\end{tabular}                                                                            & No & Yes                                                                                     \\ \hline
\cite{elsherif2019energy}                           &  \begin{tabular}[c]{@{}l@{}}Markov decision process based \\ algorithm to control the ON/OFF\\ switching of BSs.\end{tabular}                                                               & No                                             &  Yes                                                                                     \\ \hline
\cite{yu2016minimizing}                             & \begin{tabular}[c]{@{}l@{}}Adjusting transmission power \\ and considering the state transitions \\ of BSs to minimize power and \\ switching energy cost of BSs.\end{tabular} & No                                             & Yes                                                                                     \\ \hline
\cite{tang2017energy}                               & \begin{tabular}[c]{@{}l@{}}Joint user association, clustering,\\ and on/off switching strategies.\end{tabular}                                                              & No                                             & Yes                                                                                     \\ \hline
\cite{liu2015small}                                 & \begin{tabular}[c]{@{}l@{}}energy-efficient maximization \\ problem-based on coverage probability\\  under either a random sleeping \\ policy or a strategic sleeping policy\end{tabular}   & No                                             &  Yes                                                                                     \\ \hline
\end{tabular}
\end{table}

\subsection{MDP based Approach}
In \cite{ye2019drag, wu2021deep} authors formulate the \ac{BS} on/off switching problem into a Markov Decision Process that can be solved by Actor Critic (AC) reinforcement learning methods. In \cite{ye2019drag} to avoid prohibitively high computational and storage costs of conventional tabular-based approaches, authors proposed to use deep neural networks to approximate the policy and value functions in the AC approach. While in \cite{wu2021deep}, authors determine the \ac{BS} on/off modes while meeting lower energy consumption and satisfactory Quality of Service (QoS) requirements jointly using reinforcement learning methods.

Most of the earlier studies were directed toward combining optimization methods for BS on/off procedures by considering the user association with the \ac{BS}, data traffic at the \ac{BS}, etc. However, most of the aforementioned studies have primarily focused on mathematical simulations in designing the optimization algorithms, as highlighted in Table~\ref{tab:rel_work}. Network simulators like ns-3 offer several advantages over mathematical simulations when it comes to modeling and analyzing complex network systems. The advantages are detailed below. 
\begin{enumerate}
    \item \textbf{Realistic models:} ns-3 use realistic network models that are closer to the actual behavior of the network systems. These models incorporate various network protocols, traffic patterns, and network topologies that closely resemble the real world. On the other hand, mathematical simulations often rely on simplified assumptions and idealized models that may not fully capture the complexity of real-world networks.

    \item \textbf{Dynamic behavior:} ns-3 can capture the dynamic behavior of the network over time, which is difficult to achieve with mathematical simulations. For example, ns-3 can model interactions between network protocols and devices and how network traffic patterns change over time.
    
    \item \textbf{Experimentation:} ns-3 allows researchers to conduct experiments in a controlled environment, where they can test various network scenarios and configurations without affecting the real network. This is especially important for testing new protocols, applications, and technologies that may have unpredictable consequences if deployed directly on the real network.
    \item  \textbf{Scalability:} ns-3 simulations can scale to simulate large and complex networks, which is difficult to achieve with mathematical simulations. This allows researchers to study the behavior of the network under different scenarios and configurations and to identify potential problems or bottlenecks that may arise as the network grows in size and complexity.
\end{enumerate}

In summary, network simulators like ns-3 offer several advantages over mathematical simulations when it comes to modeling and analyzing complex network systems. Thus in this work, we have developed the \ac{BS} switching-off strategy in a pure network simulator, ns-3, instead of directing it to a mathematical simulation. The research community can use the developed module to implement different energy optimization schemes and test their performance.
\section{System Architecture}~\label{sec:sys_arcgh}
Since ns3-mmWave~\cite{mezzavilla2018end} lacks a well-defined \ac{UE} and \ac{BS} energy module for implementation and testing of the energy-efficient algorithms in this work we first add the UE and BS energy consumption model. Next, we discuss the implementation details of \ourmethod{} in switching off the BS based on the network load and data traffic.  
\subsection{Energy Model for the \ac{UE}}
According to the 3GPP specification 38.840~\cite{38840}, the \ac{RRC} state machine for 5G NR consists of three states: \ac{RRC} CONNECTED, RRC IDLE, and RRC INACTIVE. The level of energy consumption at UEs is determined depending on whatever state the \ac{RRC} is currently in ~\cite{patriciello2019e2e}. RRC CONNECTED indicates an active data transfer occurs between connected \ac{UE} and the \ac{BS}. When a user is connected to the \ac{BS} but no active data transmission occurs, the RRC state is IDLE. The state of the \ac{UE} between two active data transfers is reflected by the RRC INACTIVE state~\cite{polese2017improved}.

This extra planned state of RRC INACTIVE is not included in ns3 mm-Wave, which already incorporates RRC CONNECTED and RRC IDLE. To put it into action, we must map the RRC INACTIVE state to the UE PHY state. IDLE, Receive Control (RX CTRL), Receive Data (RX DATA), and Transmit (TX) are the four categories that are used to classify the PHY state of UEs~\cite{sen2021ns3}. When the UE is in the IDLE state, it indicates that there is no data transfer occurring and that the device is in a microsleep state~\cite{mezzavilla2018end}. The value of RX CTRL suggests the state of the control information exchange between the \ac{UE} and the \ac{BS}. The RX DATA state represents the data transmission state, and TX denotes the state of the up-link data transfer from the \ac{UE} to the \ac{BS}.

Using the protocol specified by 3GPP specification 38.840~\cite{38840}, the power consumption of each UE state is determined relative to the Deep sleep state, as shown in \cite[Table 18]{38840}.

Considering all the four PHY states of UE, energy consumption can be computed as $E_{UE} = \sum_{s \in S}{(P_s \times t_s)}$, where $s \in$ \\ ${\{\text{IDLE, RX CTRL, RX DATA, TX}\}}$, $E_{UE}$ means total energy consumption at UEs, $P_s$ total energy consumption for the corresponding PHY state, $P_{\text{IDLE}} = 45 mW $, $P_{\text{RX CTRL}} = 175 mW$, $P_{\text{TX}} = 350 mW$ and $P_{\text{RX DATA}} = 350mW $ from \cite{sen2021ns3}.

To compute the dwell time ($t_s$), we use the trace sink \texttt{stateChange} callback function, which gets triggered every time there is a PHY state change. The device notifies the energy source of its energy consumption, and if the energy is exhausted, the energy source alerts all related device energy models (refer to \cite[Fig. 2]{sen2021ns3}).
\subsection{\ac{BS} Energy Module}
Similarly to \acp{UE}, the energy consumption of \ac{BS} is evaluated by considering all four PHY states' energy consumption together. We use the energy model developed in \cite{sen2022ns3} in this work. This work has profiled the power consumption values in~\cite{debaillie2015flexible}. However, the developed energy model only considers the four PHY states: IDLE, RX CTRL, RX DATA, and TX. In this work, we add another PHY state called DEEP SLEEP state. The assumption is that if the \ac{BS} is IDLE for a more extended period (1s in this case), it enters the DEEP SLEEP state, which has even lower power consumption than the IDLE state. Keeping the hyperparameters detailed in \cite[Table V]{debaillie2015flexible} for $4 \times 4$ antenna, we compute power consumption values of individual PHY states as done in~\cite{sen2022ns3}. However, the only modification here is adding another state, DEEP SLEEP. In our simulations, we observe the dwell time for the IDLE state is $\approx100\mu s$. Thus according to \cite[Table V]{debaillie2015flexible} for the IDLE state, we keep the power consumption at $86.3$ watts while the DEEP SLEEP state (with dwell time $\approx 1s$) is kept at 6.2 watts (see \cite[Table V]{debaillie2015flexible}). The power values for the rest PHY state kept similar to~\cite{sen2022ns3}. These power values are shown in Table~\ref{tab:power_BS}.  

\begin{table}[h]
\scriptsize
    \centering
    \caption{Power Consumption in \ac{BS} PHY states}
    \label{tab:power_BS}
\begin{tabular}{|c|c|c|c|c|c|}
 \hline
 \textbf{State} &  $P_{\text{RX CTRL}}$ & $P_{\text{RX DATA}}$  & $P_{\text{TX}}$ & $P_{\text{IDLE}}$ & $P_{\text{DEEP SLEEP}}$\\ \hline
 \textbf{Power} & 138.9 & 138.9 & 742.2 & 86.3 & 6.2 \\ \hline
    \end{tabular}
\end{table}
Thus the total power consumption for \ac{BS} is going to be $E_{BS} = \sum_{s \in S}{(P_s \times t_s)}$ where $s \in $ \\ $\{{\text{IDLE, RX CTRL, RX DATA, TX, DEEP SLEEP}\}}$, $E_{BS}$ means total energy consumption at BSs, $P_s$ is the total power consumption for the corresponding state of PHY.

Similar to \ac{UE}, to compute the dwell time, we use the trace sink \texttt{stateChange} callback function present at the \texttt{mmWaveSpectrumPhy} class of ns-3. It gets triggered whenever the \ac{BS} changes its PHY state. The device notifies the energy source of its energy consumption, and if the energy is exhausted, the energy source alerts all related device energy models (refer to \cite[Fig. 1]{sen2022ns3}).

\subsection{Implementing \ourmethod{}}
\begin{figure}[h]
    \centering
    \includegraphics[width=0.4\textwidth]{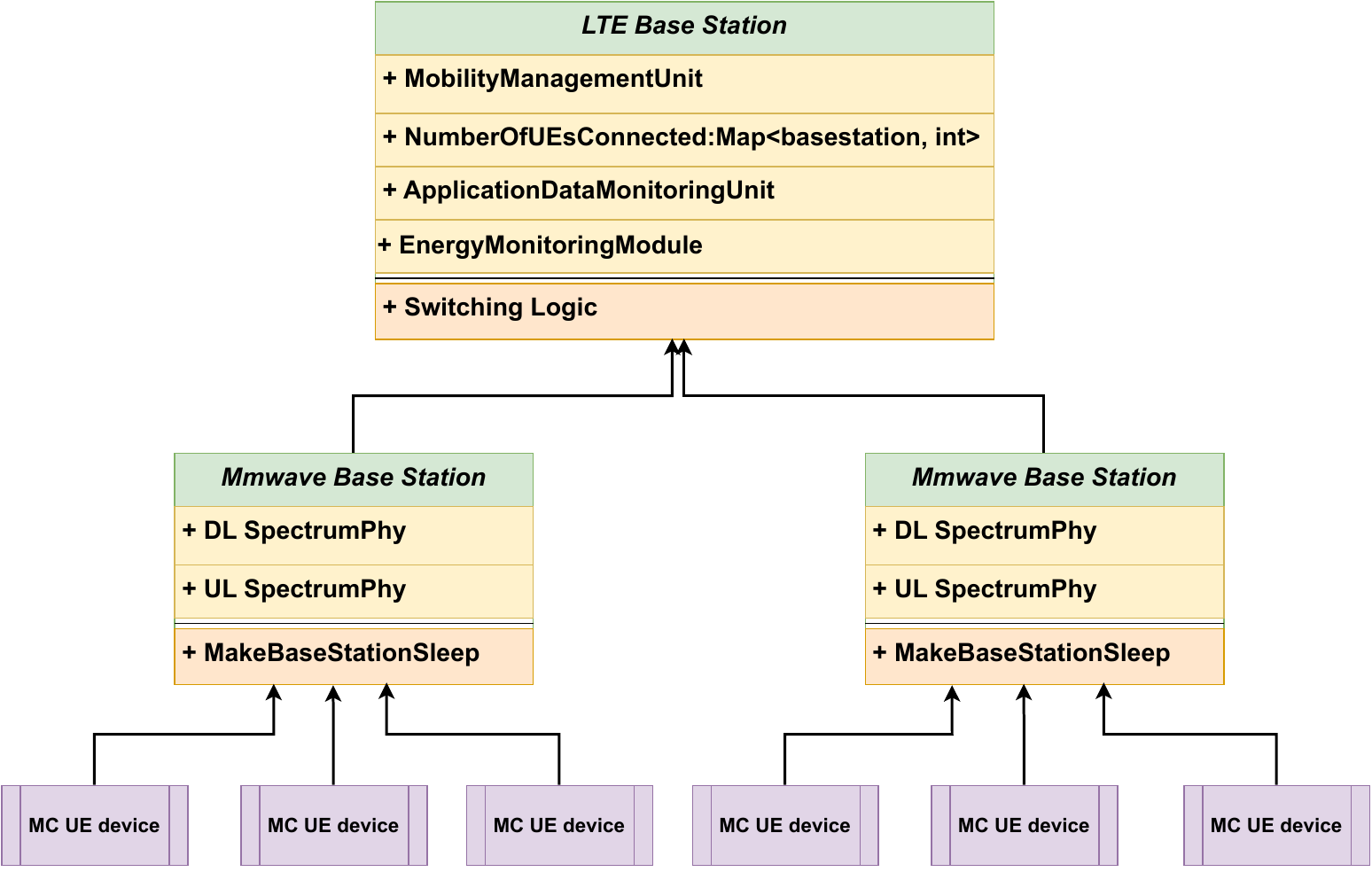}
    \caption{\ourmethod{}: Flow Diagram}
    \label{fig:smartMMEFlow}
    \vspace{-0.5cm}
\end{figure}

The implementation of \ourmethod{} involves the development of a module that resides in the LTE eNB. This module gathers all control channel information of the gNBs connected to the LTE \ac{BS} via the X2 link. The control information exchanged includes data such as the number of UEs connected to each gNB, the PHY state of the connected UEs to understand whether the UEs are IDLE or having active data transfer with the \ac{BS}, etc. The purpose of sharing this control information with the serving LTE is to analyze if any connected gNB is IDLE or connected to UEs without data transfer needs. To implement this information gathering, we leverage the standard callbacks feature of the ns-3 module. These callbacks are fired whenever there are connection establishments to some \ac{BS} or if the \ac{UE}'s are IDLE for more than $1s$ (See \figurename~\ref{fig:smartMMEFlow}). These callback functions are trace linked at the \ourmethod{} to keep track of the total number of connected UEs to a particular \ac{BS}.

When handover or connection establishment happens between the UE and the \ac{BS} using these callbacks, control channel information is shared at the \ourmethod{}. Thus the map of the number of UEs connected to each BS is updated every time connection-oriented callbacks are fired. Once the information is gathered at the Smart MME, a defined algorithm (See Algorithm~\ref{algo:switching_logic}) is used for switching ON/OFF target \ac{BS}s for energy conservation. The detail follows.

The Simulation involves a network model consisting of an LTE \ac{BS} and mmWave \ac{BS}s. The UEs are connected to the mmWave \ac{BS}s for data transfer or receive, and all UEs support \ac{NSA} dual connectivity. The \ourmethod{} node hosted in the LTE \ac{BS} has all the necessary information about each mmWave \ac{BS}, including the number of UEs connected to each mmWave \ac{BS}, the idle time of each mmWave \ac{BS}, etc.

\begin{algorithm}
\scriptsize
	\caption{\ourmethod{}: ON/OFF Switching Logic}\label{algo:switching_logic} 
	\begin{algorithmic}[1]
            \State Initialize lteBaseStation, mmWaveBaseStations, UEs
            \State Connect UEs to NearestBS
		\For {$each$ mmWaveBS $\in$ mmWaveBaseStations}
                \If{mmWaveBS.isNotConnectedToAnyUE()}
                    \State mmWaveBS.switchOff() \label{algo:swicth_off_idle}
                \ElsIf{mmWaveBS.idleTime() $ \ge 1$ sec}  
                    \State mmWaveBS.switchOff() \label{algo:swicth_off_idle_UE}
                    \EndIf
                \State connUEs $\leftarrow$ findUEs(mmWaveBS)
                \For{$each$ u $\in$ connUEs}
                    \State nearestBS $\leftarrow$ NearestBS(u) 
                    \If{nearestBS $\neq$ mmWaveBS}
                    \State mmWaveBS.disconnectUE(UE)
                    \State nearestBS.connectUE(u)
                    \ElsIf{mmWaveBS.isOff() and NearestBS(u) = mmWaveBS} 
                        \State mmWaveBS.switchOn()\label{algo:swicth_on_neighbor}
                \EndIf
                \EndFor
            \EndFor
	\end{algorithmic} 
\end{algorithm}

The switching on/off logic performs the following tasks to manage the mmWave \ac{BS}s and UEs:

\begin{enumerate}
    \item \textbf{Switch off IDLE mmWave \ac{BS}:} \ourmethod{} checks the IDLE time of each mmWave \ac{BS} and switches off those that are not connected to any UEs.
    \item \textbf{Switch off mmWave \ac{BS} with no data transfer for more than 1 seconds:} For each mmWave BS \ourmethod{} checks if any \ac{BS} is connected to multiple UEs but the UEs are IDLE for more than 1 second, i.e., there is no data transfer it switches off the \ac{BS}.
    \item \textbf{Switch on mmWave \ac{BS} if nearest to UE:} \ourmethod{} checks if a UE enters in the coverage region of previously switched off \ac{BS}, and it then switches on the \ac{BS}.
\end{enumerate}

These instructions are intended to optimize network performance and resource usage by balancing the load across the network and improving coverage and connectivity for the UEs.

\section{Evaluation}~\label{sec:eval}
\begin{figure*}
    \centering
    \subfigure[Default Scenario]{
    \includegraphics[height=3.03cm,width=0.48\columnwidth]{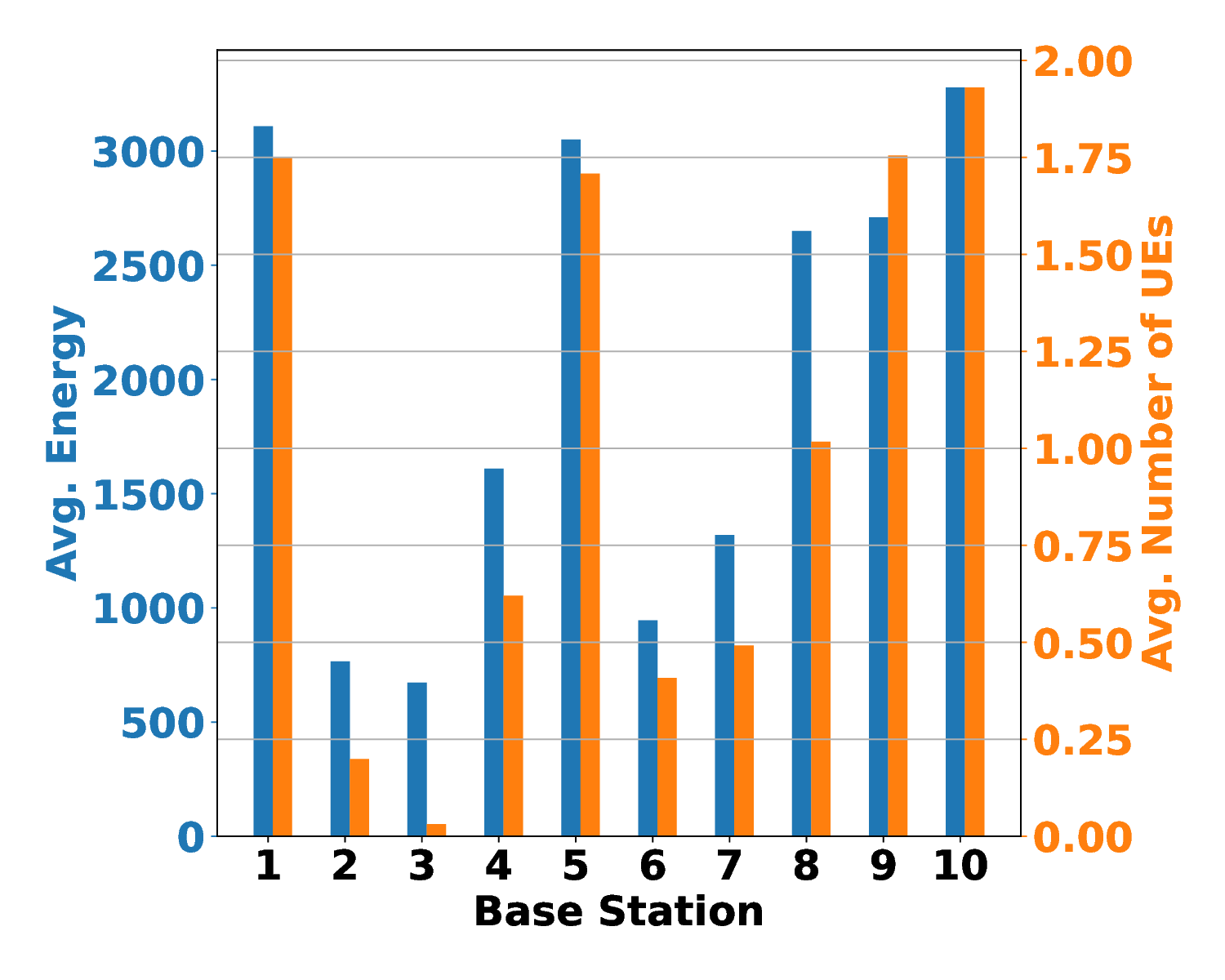}\label{fig:default}
    }\hfil
    \subfigure[Random Switching Off]{
    \includegraphics[height=3.03cm,width=0.48\columnwidth]{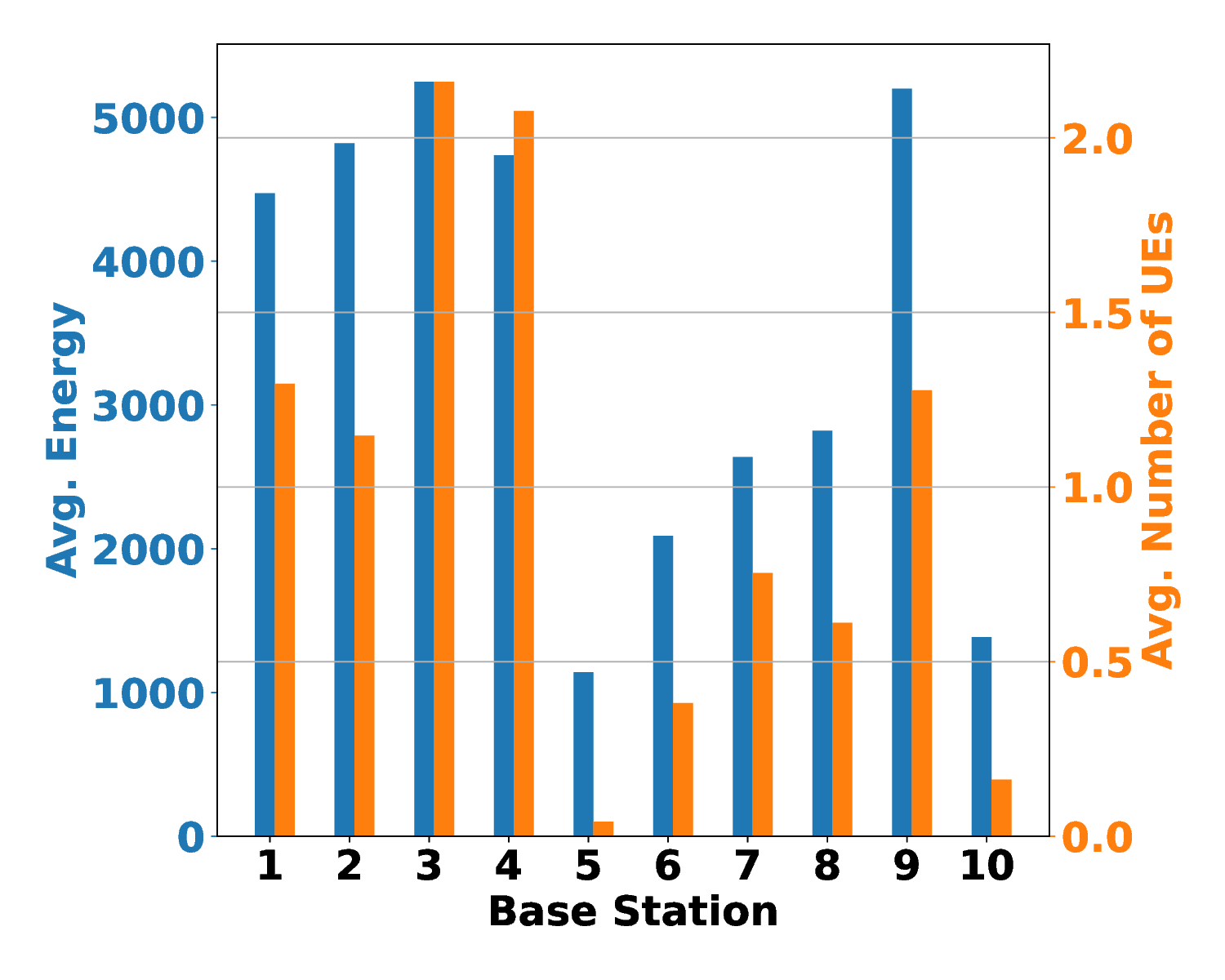}\label{fig:RandomSwitchingOff}
    }\hfil
   \subfigure[\# of UE aware]{
    \includegraphics[height=3.03cm,width=0.48\columnwidth]{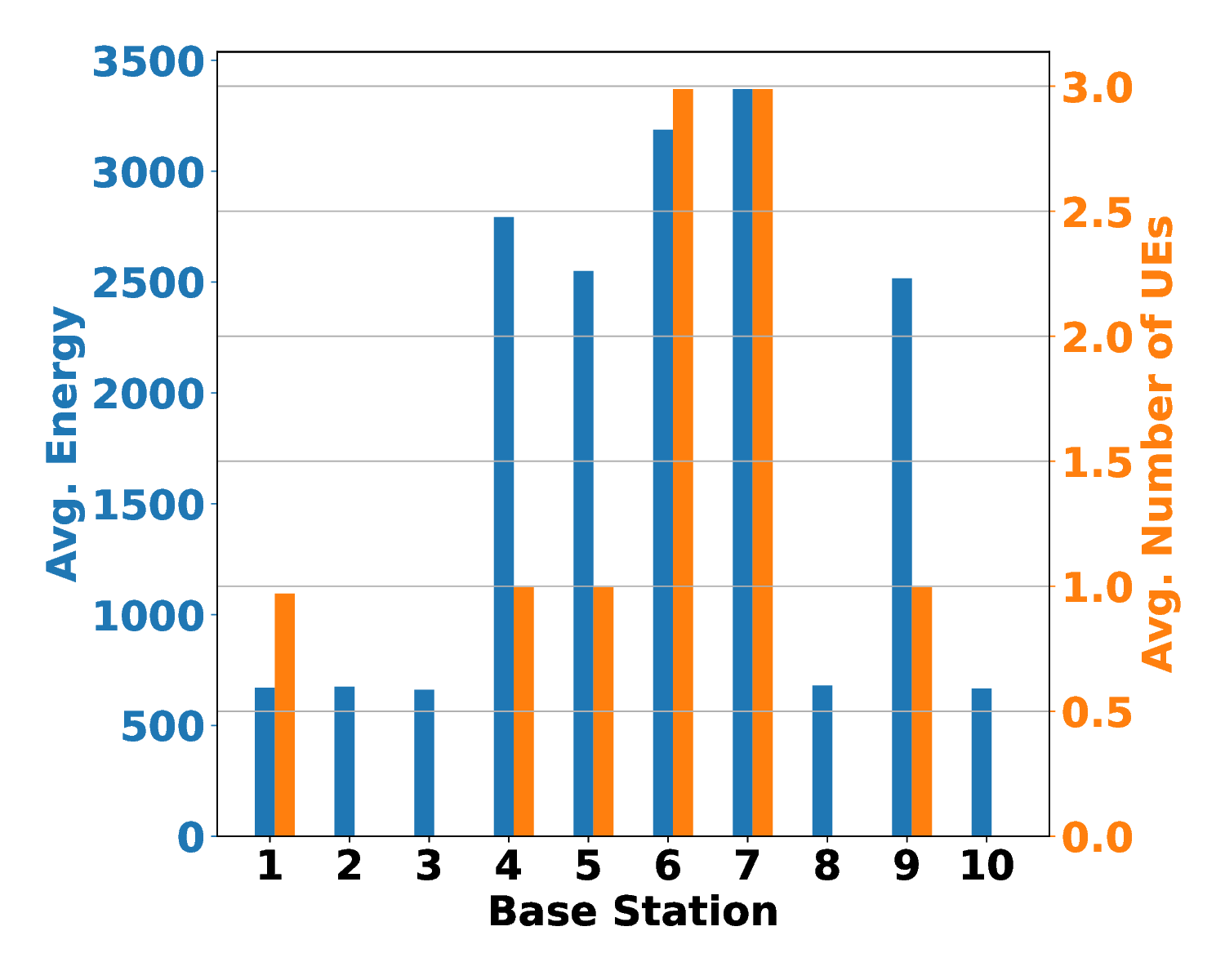}\label{fig:ue_aware}
    }\hfil
    \subfigure[Application aware]{
    \includegraphics[height=3.03cm,width=0.48\columnwidth]{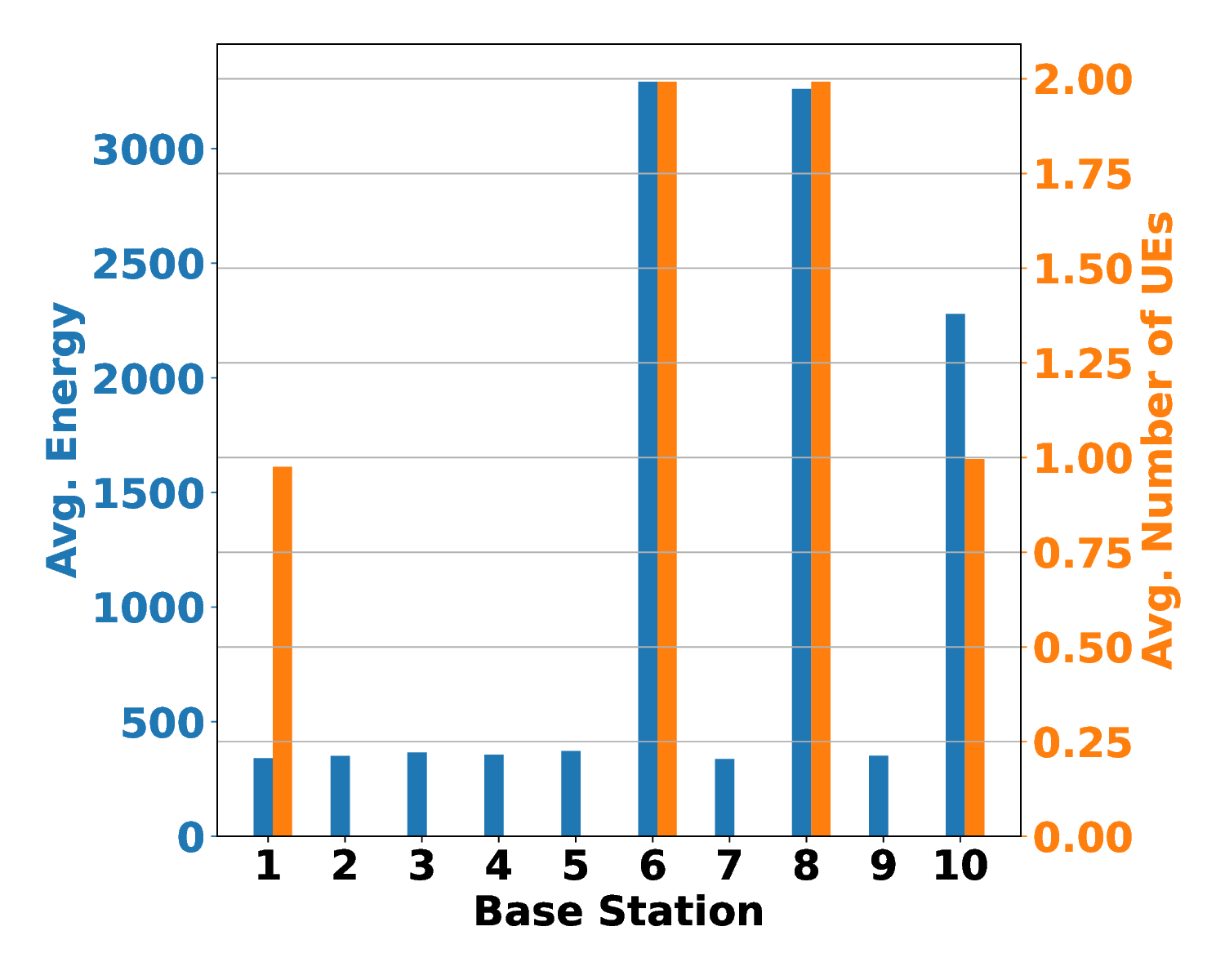}\label{fig:dataAware}
    }
    \caption{Energy Consumption and average number of UEs connected across different \ac{BS}s}
    \label{fig:energy_avg_ue}
    \vspace{-0.5cm}
\end{figure*} 

The simulation parameters are tabulated in Table~\ref{tab:sim_parameters}. We have taken four different simulation scenarios: (i) \textbf{Basic Simulation Scenario} where we keep $10$ UEs and $10$ \ac{BS}s, (ii)  \textbf{Random Switching Off} where we randomly switch off the \ac{BS}, (iii) \textbf{UE Connectivity Aware} where \ourmethod{} switches off the \ac{BS} if no UEs are connected to it, (iv) \textbf{UE Connectivity and Data Aware} where \ourmethod{} switches off the \ac{BS} if no UEs are connected or if the connected \acp{UE} are idle.
    \begin{table}
   \scriptsize 
        \centering
        \rule{0pt}{2em}
        \caption{Simulation Parameters}
        \label{tab:sim_parameters}
        \begin{tabular}{|p{5cm}|p{3cm}|} 
         \hline
        \textbf{Parameter Description} & \textbf{Value} \\
         \hline
         \hline
          Bandwidth of mmwave gNBs/LTE eNB & 1 GHz/20 MHZ \\
              \hline
        Carrier frequency mmwave/LTE & 28 GHz/2.1GHz  \\     \hline
    
    
         Bandwidth of the LTE eNB & 20 MHz  \\     \hline

        MIMO array size gNB/UE & $8 \times 8/4 \times 4$   \\     \hline
    
         Number of gNB/eNB &10 /1 \\     \hline
    
         Number of \acp{UE} & 10   \\     \hline
    
     UE speed & 5 m/s\\     \hline
    UE Application & DL UDP Socket App. \\ \hline
    UE Mobility Model & Random Walk 2d \\ \hline
    BS Mobility Model & Constant Position \\ \hline
    Path loss Model & 3GPP Umi Street Canyon \\ \hline
    Channel Condition Model & Buildings Channel \\ \hline
    Area of Simulation & 1 km $\times$ 1 km \\ \hline
    Simulation Time & 11.24 sec \\ \hline
         \end{tabular}
         \rule[-1em]{0pt}{1em}
         \label{table:2}
             \vspace*{-0.5cm}
     \end{table} 
     
\begin{figure}
    \centering
    \includegraphics[width=0.35\textwidth]{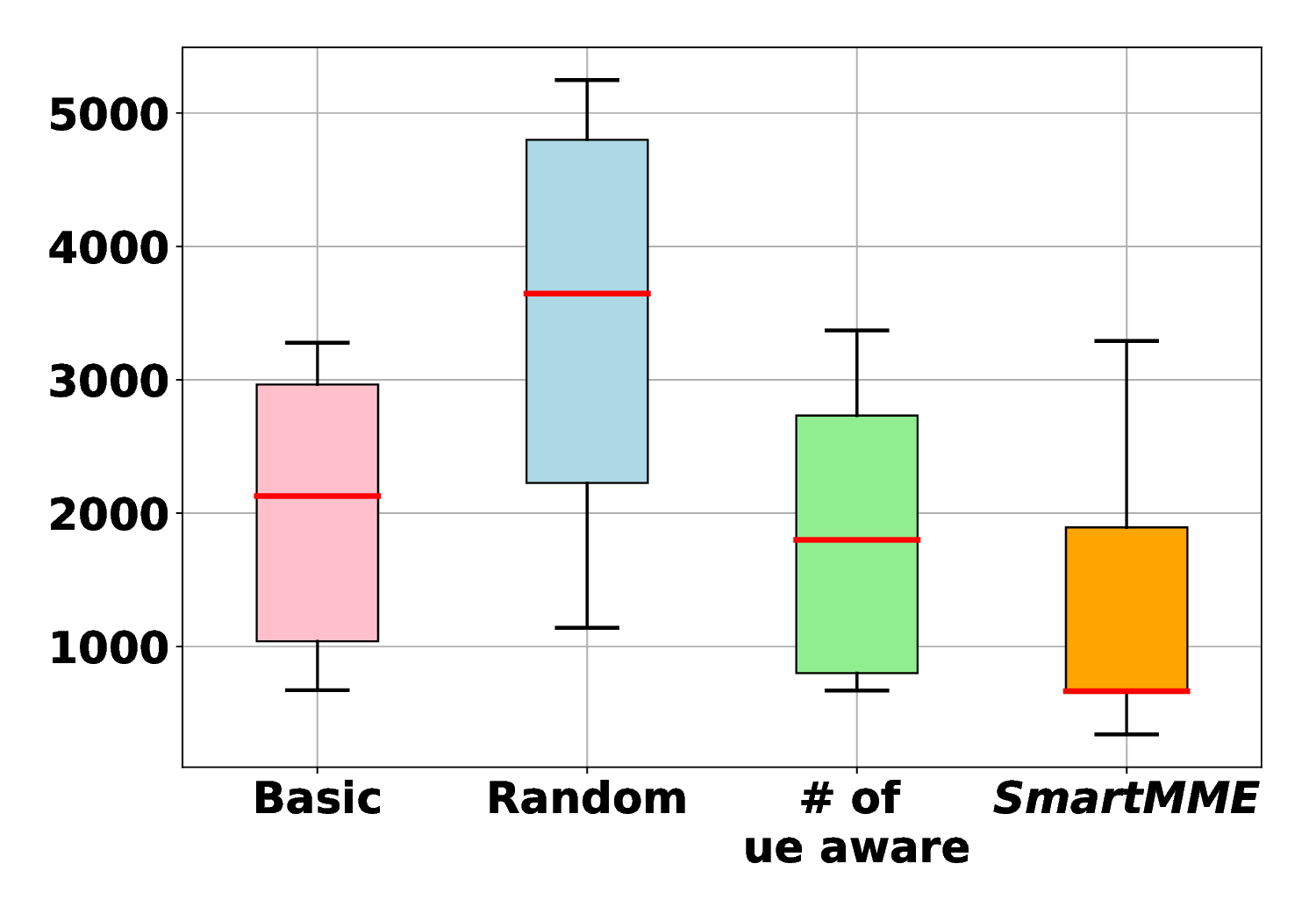}
    \caption{Energy Consumption across different scenarios}
    \label{fig:overall_E}
    \vspace{-0.5cm}
\end{figure}


\subsection{Basic Simulation Scenario}
In the basic simulation scenario, we simulate without any switching-off strategy for 11.24 seconds. Ten UEs and ten gNBs were deployed in a $1000 \times 1000$ rectangular region, randomly in a uniform distribution. \figurename~\ref{fig:default} shows the energy consumption per \ac{BS} varying with the number of UEs connected to it. The general trend is that the more the number of UEs, the greater the energy consumption at that \ac{BS}.

\subsection{Random Switching Off}
This experiment ran the basic simulation with randomly switching off the \ac{BS}s. Here the \ac{BS}s are scheduled to switch off for a period of $2$ seconds in a simulation of $11.24$ seconds. For each \ac{BS}, a random integer less than $9$ is generated using the \texttt{rand()} function of the standard \texttt{C++} library, say $X$, and the \ac{BS} is scheduled to sleep from $X$ seconds to $X+2$ seconds. \figurename~\ref{fig:RandomSwitchingOff} shows the energy consumption per \ac{BS} varying with the no. of UEs connected to it. The energy consumption is higher than basic simulation because randomly switching off a \ac{BS} entails a lot of overhead of handovers to ensure the QoE.

\subsection{UE Connectivity Aware Switching Off}
In this experiment, with the help of our \ourmethod{} node, we simulate by switching off a \ac{BS} if there is no UE connected to it. \ourmethod{} keeps track of the number of UEs connected to a \ac{BS} in the simulation, which is updated upon every connection establishment or handover. \figurename~\ref{fig:ue_aware} shows the energy consumption per \ac{BS} varying with the number of UEs connected to it. It can be seen that there are a few \ac{BS}s with deficient energy consumption where no UEs are connected. In this case, the \ac{BS} enters the deep sleep state, and thus the energy consumption is deficient compared to other scenarios.

\subsection{UE Connectivity and Data Application Aware Switching Off}
Finally, in this experiment, we run the simulation where \ourmethod{} works in the full-fledged mode, where it collects both the information of the number of UEs connected to a \ac{BS} and checks if all the connected \acp{UE} in a particular \ac{BS} is IDLE for more than $1$ second or not. Here we deploy a single \ac{BS} (say $BS_1$) with 4 UEs deployed close to it. These UEs do not run any application. We deploy the rest 9 \acp{BS} too far away from the initial setting and deploy the rest of the $6$ UEs around these \ac{BS}s. These UEs are running an application over them. The UEs and \ac{BS}s were deployed with a \textit{ConstantMobilityModel} to avoid handovers. Since the UEs connected to $BS_1$ runs no application, \ourmethod{} switches off the $BS_1$ as there is no data flow towards or from it. Illustrated in \figurename~\ref{fig:dataAware} is the energy consumption per \ac{BS}, delineating its fluctuations concerning the number of connected UEs. Notably, the most efficient energy consumption is attributed to $BS_1$, which remains switched off. Conversely, the remaining \acp{BS} operate in regular mode, dynamically adjusting their energy consumption in response to the varying number of connected UEs.

The energy consumption distribution across the four scenarios is depicted in \figurename~\ref{fig:overall_E}. Notably, \ourmethod{} demonstrates superior energy efficiency when operating in its comprehensive mode, intelligently considering both the number of connected UEs and data traffic. In contrast, the UE Connectivity-aware scenario exhibits slightly higher power consumption due to the inability to switch off \acp{BS} with IDLE UEs. Conversely, the Random Switching off the \ac{BS} scenario records the highest energy consumption, primarily attributed to frequent handovers and connection re-establishment. This underscores \ourmethod{}'s adeptness in harnessing network and data traffic insights to strategically power down \acp{BS}, thus significantly enhancing overall energy efficiency across the network.

\section{Conclusion}\label{sec:conclusion}
Our work introduces \ourmethod{}, an adaptive \ac{BS} on/off switching system finely attuned to UE and data traffic dynamics. The open-source availability of this pivotal module extends a valuable opportunity to the wider research community, enabling comprehensive evaluations and refinements of diverse power-saving strategies within the network simulation domain of the ns-3 platform. Through rigorous evaluation, we've demonstrated its significant impact on enhancing energy efficiency by intelligently responding to evolving traffic dynamics within specific \ac{BS} coverage areas. In future work, we aim to bolster the module's efficacy by incorporating real-world mobility traces from platforms like Simulation of Urban MObility (SUMO), fortifying its adaptability in realistic urban scenarios. Additionally, our ongoing focus involves evaluating alternative switching strategies from existing literature to optimize energy efficiency in dynamic network environments.
\bibliographystyle{IEEEtran}

\bibliography{main.bib}
\balance
\begin{acronym}
	\acro{2G}{2$^\text{nd}$ Generation}
	\acro{3G}{3$^\text{rd}$ Generation}
	\acro{4G}{4$^\text{th}$ Generation}
	\acro{5G}{5$^\text{th}$ Generation}
	\acro{A3C}{Actor-Critic}
	\acro{ABR}{adaptive bitrate}
	\acro{BS}{Base Station}
	\acro{CDN}{Content Distribution Network}
	\acro{DASH}{Dynamic Adaptive Streaming over HTTP}
	\acro{DL}{deep learning}
	\acro{DRX}{Discontinuous Reception}
	\acro{EDGE}{Enhanced Data Rates for \ac{GSM} Evolution.}
			\acro{gNB}{general NodeB}
	\acro{eNB}{evolved NodeB}
	\acro{GSM}{Global System for Mobile}
	\acro{FL}{Federated Learning}
	\acro{TL}{Transfer Learning}
	\acro{HD}{High Definition}
	\acro{HSPA}{High Speed Packet Access}
	\acro{LSTM}{Long Short Term Memory}
	\acro{LTE}{Long Term Evolution}
	\acro{ML}{machine learning}
	\acro{MTL}{Multi-Task Learning}
	\acro{MCS}{Modulation and Coding Scheme}
	\acro{NSA}{Non-Standalone}
	\acro{HVPM}{High voltage Power Monitor}
	\acro{QoS}{Quality of Service}
	\acro{QoE}{Quality of Experience}
	\acro{RF}{Random Forest}
	\acro{RFL}{Random Forest}
	\acro{RL}{Reinforcement Learning}
	\acro{RRC}{Radio Resource Control}
	\acro{RSSI}{Received Signal Strength Indicator}
	\acro{RSRP}{Reference Signal Received Power}
	\acro{RSRQ}{Reference Signal Received Quality}
	\acro{SINR}{signal-to-interference-plus-noise-ratio}
	\acro{SNR}{signal-to-noise-ratio}
	\acro{UE}{User Equipment}
	\acro{UHD}{Ultra HD}
	\acro{VoLTE}{Voice over LTE}
	\acro{RNN}{Recurrent Neural Network}
	\acro{WiFi}{Wireless Fidelity}
	\acro{ARIMA}{Auto Regressive Integrated Moving Average}
	\acro{ML}{Machine Learning}
	\acro{NR}{New Radio}
	\acro{Non-IID} {Independent and Identically Distributed}
	\acro{TP}{Throughput Prediction}
	\acro{CQI}{Channel Quality Indicator}
	\acro{CTFL}{Cross-Technology Federated Learning}
        \acro{NAA}{Network Aware Application}
        \acro{SVR}{Support Vector Regression}
        \acro{MLP}{Multilayer Perceptrons}
        \acro{UDN}{Ultra Dense Networks}
\end{acronym}

\end{document}